\documentstyle[epsf,onecolumn]{mn}

\newcommand{\eck}[1]{\left[#1\right]}
\newcommand{\rund}[1]{\left(#1\right)}

\begin{document}

\title[Aperture Multipole Moments]
      {Aperture Multipole Moments from Weak Gravitational Lensing}

\author[Peter Schneider \& Matthias Bartelmann]
       {Peter Schneider and Matthias Bartelmann\\
	Max-Planck-Institut f\"ur Astrophysik, Postfach 1523,
	D-85740 Garching, Germany}

\maketitle

\begin{abstract}
The projected mass of a gravitational lens inside (circular) apertures
can be derived from the measured shear inside an annulus which is
caused by the tidal field of the deflecting mass distribution. Here we
show that also the multipoles of the two-dimensional mass distribution
can be derived from the shear in annuli. We derive several expressions
for these mass multipole moments in terms of the shear, which allow
large flexibility in the choice of a radial weight function. In
contrast to determining multipole moments from weak-lensing mass
reconstructions, this approach allows to quantify the signal-to-noise
ratio of the multipole moments directly from the observed galaxy
ellipticities, and thus to estimate the significance of the multipole
detection. Radial weight functions can therefore be chosen such as to
optimize the significance of the detection given an assumed radial
mass profile. Application of our formulae to numerically simulated
clusters demonstrates that the quadrupole moment of realistic cluster
models can be detected with high signal-to-noise ratio $S/N$; in
$\simeq85$ per cent of the simulated cluster fields $S/N\ga3$. We also
show that the shear inside a circular annulus determines multipole
moments inside and outside the annulus. This is relevant for clusters
whose central region is too bright to allow the observation of the
shear of background galaxies, or which extend beyond the CCD.  We also
generalize the aperture mass equation to the case of `radial' weight
functions which are constant on arbitrarily-shaped curves which are
not necessarily self-similar.
\end{abstract}

\section{Introduction}

As was shown by Kaiser \& Squires (1993), the observable image
distortion of high-redshift galaxies by an intervening mass
concentration acting as a gravitational lens can be used to
reconstruct the two-dimensional mass map of the deflector. The local
mean image distortion provides an estimate for the local tidal
gravitational field, from which the underlying surface mass density
can be obtained. Since then, the original inversion method has been
generalized in several ways (see, e.g., Schneider 1996a for a recent
review and references). The mass maps obtained in this way are noisy,
and the significance of individual features is not easily
assessed. Kaiser (1995) proposed to use a measure for the mass
convolved with a compensated top-hat filter. This so-called
$\zeta$-statistics, or aperture mass, has the advantage that the
statistical properties of this mass estimate can be easily obtained
from the observations. This statistics was then modified to allow for
more general radial weight functions (Kaiser et al. 1994), and
proposed by Schneider (1996b) as a means to search for `dark' mass
halos on wide-field images.

Recently, Wilson, Cole, \& Frenk (1996) have investigated the
possibility of studying the asymmetry of galaxy clusters with
weak-lensing methods. As is well known (e.g., Richstone, Loeb, \&
Turner 1992; Bartelmann, Ehlers, \& Schneider 1993), the amount of
asymmetry in clusters depends quite strongly on the cosmological
model, since it signifies the degree of relaxation, and thus age, of
the clusters. High-density models predict a much later cluster
formation epoch, and thus more asymmetric clusters, than low-density
models of the universe. Wilson et al.\ (1996) applied the Kaiser \&
Squires (1993) cluster mass reconstruction algorithm to a sample of
numerically generated cluster models and investigated the quadrupole
moment of the {\em area} enclosed by a selected isodensity
contour. Doing so, they showed that this measure indeed discriminates
well between high- and low-$\Omega$ universes if applied to a fairly
small number of massive clusters.

Here we derive mass multipole moments in circular apertures in terms
of the shear in an annulus. As in the case of aperture masses, the
chief advantage of aperture measures for the multipole moments is its
direct relation to observable quantities, which allows to
unambiguously quantify the accuracy, and thus the significance, of the
result. Multipole moments can also be determined from mass maps
derived from cluster mass reconstructions (e.g., Kaiser \& Squires
1993; Seitz \& Schneider 1995; Bartelmann et al.\ 1996; Wilson et al.\
1996). Apart from being technically more involved, the inference in
this case is indirect because multipoles are determined from the
reconstructed mass maps rather than directly from the shear data, it
does not allow to give simple error estimates, and it depends on
artificial parameters like a smoothing length.

In contrast to the aperture mass, the weight functions (filters) do
not need to be compensated. With two different methods, the aperture
multipoles in terms of the shear field are derived in Sect.\ 2,
leading to two apparently different expressions. It is shown in
Appendix A that these two expressions are equivalent, and they can be
suitably combined to allow for a large flexibility in the choice of
the weight function. Several examples for this are presented. Alike
the aperture mass, the mass multipoles inside a circle can be derived
from the shear in an annulus outside that circle. In particular, as
will be shown explicitly in Appendix B, the shear in a circular
annulus is determined by the mass in this annulus, and the multipole
moments of the mass inside and outside the annulus. We then
investigate in Sect.\ 3 the signal-to-noise statistics of the aperture
multipoles, using as a simple example a quasi-elliptical mass
distribution. In Sect.\ 4 we present a preliminary analysis of the
quadrupole moments of numerically generated cluster models to
demonstrate the applicability of our method to realistic clusters. As
discussed in Sect.\ 5, the application to a larger sample of cluster
models drawn from different cosmological simulations provides an
alternative approach to the discrimination between different
cosmological scenarios.

The methods used for deriving the aperture multipoles are then used in
Appendix C to generalize the aperture mass to weight functions which
are constant on a set of nested arbitrarily-shaped curves which are
not necessarily self-similar.
 
\section{Aperture multipoles}

\subsection{Definitions}

Consider a mass distribution, e.g., a cluster of galaxies, at redshift
$z_{\rm d}$, with surface mass density $\Sigma(\vec x)$. The
dimensionless surface mass density $\kappa(\vec x)$ is defined as
usual in gravitational lensing,
\begin{equation}
  \kappa(\vec x) = {\Sigma(\vec x)\over\Sigma_{\rm cr}}\;,
\label{eq:1}
\end{equation}
with the critical density
\begin{equation}
  \Sigma_{\rm cr} = 
  {c^2 D_{\rm s}\over4\pi G\,D_{\rm d}\,D_{\rm ds}}\;,
\label{eq:2}
\end{equation}
and the $D$'s denote the angular-diameter distances between observer,
deflector and source, and between deflector and source (for notation,
see Schneider, Ehlers, \& Falco 1992). For simplicity, we will
provisionally assume that all faint galaxies are at the same redshift
$z_{\rm s}$. The generalization to sources distributed in redshift
will be discussed in Sect.\ 3.

We define the tensor of quadrupole moments of the two-dimensional mass
distribution $\kappa(\vec x)$ with respect to the point $\vec x_0$ as
\begin{equation}
  Q_{ij} = \int{\rm d}^2x\;\kappa(\vec x+\vec x_0)\,
  w\rund{|\vec x|}\,x_i\,x_j\;,
\label{eq:3}
\end{equation}
for $i,j\in\{1,2\}$, where $w(x)$ is a radial weight function.
Combining the trace-free part of this tensor into a complex number
$Q=Q_{11}-Q_{22}+2{\rm i} Q_{12}$, this can be written as
\begin{equation}
  Q = \int_0^\infty{\rm d}x\,x^3\,w(x)
  \int_0^{2\pi}{\rm d}\varphi\,{\rm e}^{2{\rm i}\varphi}\,
  \kappa(\vec x+\vec x_0)\;,
\label{eq:4}
\end{equation}
whereas the trace of $Q_{ij}$ becomes
\begin{equation}
  M := Q_{11}+Q_{22} = \int_0^\infty{\rm d}x\,x^3\,w(x)
  \int_0^{2\pi}{\rm d}\varphi\,\kappa(\vec x+\vec x_0)\;.
\label{eq:5}
\end{equation}
These equations are easily generalized for higher multipole orders. We
define the complex $n^{\rm th}$-order multipole moment by
\begin{equation}
  Q^{(n)} := \int_0^\infty{\rm d}x\,x^{n+1}\,w(x)
  \int_0^{2\pi}{\rm d}\varphi\,{\rm e}^{n{\rm i}\varphi}
  \kappa(\vec x+\vec x_0)\;,
\label{eq:6}
\end{equation}
and the `mass moments'
\begin{equation}
  M^{(n)} := \int_0^\infty{\rm d}x\,x^{n+1}\,w(x)
  \int_0^{2\pi}{\rm d}\varphi\,\kappa(\vec x+\vec x_0)\;.
\label{eq:7}
\end{equation}
Obviously, $Q^{(2)}\equiv Q$ and $M^{(2)}\equiv M$.

\subsection{Multipole moments in terms of the shear}

Let $\psi(\vec x)$ denote the deflection potential, which is related
to the surface mass density $\kappa$ through the Poisson-like equation
\begin{equation}
  \kappa = {1\over 2}\nabla^2\psi =
  {1\over2}\rund{\psi_{,11}+\psi_{,22}}\;,
\label{eq:8}
\end{equation}
where indices preceded by a comma denote partial derivatives. The two
components of the local shear $\gamma(\vec x)=\gamma_1+{\rm
i}\gamma_2$ are then given by
\begin{equation}
  \gamma_1 = {1\over2}\rund{\psi_{,11}-\psi_{,22}}\quad,\quad
  \gamma_2 = \psi_{,12}\;.
\label{eq:9}
\end{equation}
It was shown by Kaiser (1995) by suitably combining third-order
derivatives of $\psi$ that the gradient of $\kappa$ can be written in
terms of derivatives of the shear components as follows:
\begin{equation}
  \nabla\kappa=\pmatrix{\gamma_{1,1}+\gamma_{2,2}\cr
                        \gamma_{2,1}-\gamma_{1,2}}\;.
\label{eq:10}
\end{equation}
This relation can now be used to express the multipole moments defined
above in terms of the shear $\gamma$. Since the shear is directly
observable from the distortion of the images of faint background
galaxies (at least in the case of weak lensing, i.e., $\kappa\ll 1$),
the multipole moments can be written directly in terms of observables.

For later convenience, we define the {\em tangential shear
$\gamma_{\rm t}(\vec x;\vec x_0)$ and the radial shear $\gamma_{\rm
r}(\vec x;\vec x_0)$ at position $\vec x=x\,{\rm e}^{{\rm i}\varphi}$
relative to position $\vec x_0$} by
\begin{eqnarray}
  \gamma_{\rm t}(\vec x;\vec x_0) &=&
   -\eck{\gamma_1\cos(2\varphi)+\gamma_2\sin(2\varphi)} =
   -\Re\eck{\gamma(\vec x+\vec x_0)\,{\rm e}^{-2{\rm i}\varphi}}\;,
  \nonumber\\
  \gamma_{\rm r}(\vec x;\vec x_0) &=&
   -\eck{\gamma_2\cos(2\varphi)-\gamma_1\sin(2\varphi)} =
   -\Im\eck{\gamma(\vec x+\vec x_0)\,{\rm e}^{-2{\rm i}\varphi}}\;,
  \nonumber\\
\label{eq:11}
\end{eqnarray}
where $\Re(z)$ and $\Im(z)$ denote the real and imaginary part of the
complex number $z$.

\subsubsection{First derivation}

We integrate the $\varphi$-integral in (\ref{eq:6}) by parts to obtain
\begin{equation}
  Q^{(n)} = {{\rm i}\over n}\,\int_0^\infty{\rm d}x\,x^{n+1}\,w(x)\,
  \int_0^{2\pi}{\rm d}\varphi\,{\rm e}^{{\rm i}\,n\varphi}\,
  {\partial\kappa\over\partial\varphi}\;.
\label{eq:12}
\end{equation}
The derivative of $\kappa$ with respect to $\varphi$ can be
substituted after transforming (\ref{eq:10}) to polar
coordinates. Subsequently, terms containing partial derivatives with
respect to $x$ and $\varphi$ can be partially integrated with respect
to $x$ and $\varphi$, respectively. This leads to
\begin{equation}
  Q^{(n)} \equiv Q^{(n)}_\varphi =
  \int_0^\infty{\rm d}x\,x^{n+1}\,w(x)
  \int_0^{2\pi}{\rm d}\varphi\,{\rm e}^{n{\rm i}\varphi}\,
  \gamma_{\rm t}(\vec x;\vec x_0) +
  {{\rm i}\over n}\,\int_0^\infty{\rm d}x\,
  \eck{n x^{n+1} w(x)+x^{n+2}w'(x)}\,
  \int_0^{2\pi}{\rm d}\varphi\,{\rm e}^{n{\rm i}\varphi}\,
  \gamma_{\rm r}(\vec x;\vec x_0)\;,
\label{eq:13}
\end{equation}
where $w'(x)$ is the derivative of $w(x)$, and $w(x)$ has to be
continuous and piecewise differentiable. In order to get rid of the
boundary terms in the partial integrations leading to (\ref{eq:13}),
$w(x)$ has to satisfy the conditions
\begin{equation}
  |\gamma|\,x^{n+2}\,w(x)\to 0\quad
  {\rm for}\quad x\to 0\quad{\rm and}\quad x\to\infty\;.
\label{eq:14}
\end{equation}

\subsubsection{Second derivation}

Alternatively, we can integrate (\ref{eq:6}) by parts with respect to
$x$ and find
\begin{equation}
  Q^{(n)} = -\int_0^\infty{\rm d}x\,x\,W(x)
  \int_0^{2\pi}{\rm d}\varphi\,{\rm e}^{n{\rm i}\varphi}\,
  {\partial\kappa\over\partial x}\;,
\label{eq:15}
\end{equation}
where $W(x)$ is related to the weight function $w(x)$ by
\begin{equation}
  x\,W(x) = \int_0^x{\rm d} y\; y^{n+1}\,w(y)\;.
\label{eq:16}
\end{equation}
Substituting into (\ref{eq:15}) the radial derivative of $\kappa$ from
(\ref{eq:10}) transformed to polar coordinates, we obtain
\begin{equation}
  Q^{(n)} \equiv Q^{(n)}_x =
  \int_0^\infty{\rm d}x\,\eck{2 W(x)-x^{n+1}\,w(x)}\,
  \int_0^{2\pi}{\rm d}\varphi\,{\rm e}^{n{\rm i}\varphi}\,
  \gamma_{\rm t}(\vec x;\vec x_0) -
  {\rm i}\,n\,\int_0^\infty{\rm d}x\,W(x)
  \int_0^{2\pi}{\rm d}\varphi\,{\rm e}^{n{\rm i}\varphi}\,
  \gamma_{\rm r}(\vec x;\vec x_0)\;.
\label{eq:17}
\end{equation}
The boundary terms in the partial integration leading to (\ref{eq:15})
vanish if $W(x)$ satisfies
\begin{equation}
  x\,W(x)\,\kappa\to 0\quad
  {\rm for}\quad x\to 0\quad{\rm and}\quad x\to\infty\;.
\label{eq:18}
\end{equation}

We have thus derived two expressions for $Q^{(n)}$ which are of quite
different form. We prove their equivalence in Appendix A, using the
fact that the two components of the shear are not mutually
independent, but related via the underlying deflection
potential. Another way to see the equivalence is to note that the curl
of the vector field on the right-hand-side of (\ref{eq:10}) must vanish,
which is a condition not used in the derivation of either (\ref{eq:13})
or (\ref{eq:17}).

After a sequence of partial integrations similar to those used before,
the mass moments (\ref{eq:7}) become
\begin{equation}
  M^{(n)} = \int_0^\infty{\rm d}x\,\eck{2 W(x)-x^{n+1}\,w(x)}\,
  \int_0^{2\pi}{\rm d}\varphi\,\gamma_{\rm t}(\vec x;\vec x_0)\;,
\label{eq:19}
\end{equation}
which is the mass aperture equation of Kaiser et al.\ (1994; see also
Schneider 1996b).

\subsection{Aperture multipole measures}

The fact that we have two different, but equivalent, expressions for
the multipole moments $Q^{(n)}$ provides some freedom in choosing the
weight functions $w(x)$ and $W(x)$ suitably. Recall the situation for
the aperture mass measures as discussed in Schneider (1996b). There,
from shear data in a ring with $x\in[\nu R,R]$ with $\nu<1$, the
aperture mass measure $m\equiv M^{(0)}$ can be determined, provided
the weight function $w(x)$ satisfies $\int{\rm d} x\,x\,w(x)=0$. This
condition arises because the shear determines the surface mass density
only up to an additive constant, and the normalization condition on
$w$ guarantees that this unknown additive constant drops out of the
aperture mass. As an aside, we generalize the aperture mass formula to
arbitrarily shaped apertures in Appendix C.

Similarly, we assume here that shear data are given in the ring
$x\in[\nu R,R]$, where in practice the outer radius $R$ is determined
by the size of the CCD, and an inner radial distance from the cluster
center may be required if the cluster contains bright central galaxies
which outshine the images of faint background galaxies. Then,
$Q^{(n)}_\varphi$ is a {\em local} estimator for $Q^{(n)}$, in the
sense that if the weight function $w(x)$ is nonzero only in a
specified radial interval, the estimate $Q^{(n)}_\varphi$ requires
shear data only from that interval. However, it should be noted that
the boundary terms in the integrations by parts with respect to $x$
leading to (\ref{eq:13}) vanish only if $w(x)=0$ at the boundaries of
the interval.

In contrast to $Q^{(n)}_\varphi$, $Q^{(n)}_x$ is a {\em non-local}
estimate of $Q^{(n)}$, for it requires shear data from outside the
interval, unless $W(x)$ also vanishes outside the specified
interval. It follows from (\ref{eq:16}) that $W(x)=0$ outside an
interval imposes integral constraints on $w(x)$, whereas $w(x)$ is an
arbitrary function in the estimator $Q^{(n)}_\varphi$ (except that it
has to vanish at the interval boundaries).

To simplify notation in the remainder of the paper, we introduce the
abbreviations
\begin{equation}
  g_{\rm t}(x) := \int_0^{2\pi}{\rm d}\varphi\,
  {\rm e}^{n{\rm i}\varphi}\,\gamma_{\rm t}(\vec x;\vec x_0)
  \quad,\quad
  g_{\rm r}(x) := {\rm i}\int_0^{2\pi}{\rm d}\varphi\,
  {\rm e}^{n{\rm i}\varphi}\,\gamma_{\rm r}(\vec x;\vec x_0)\;.
\label{eq:20}
\end{equation}

\subsubsection{Local estimate of $Q^{(n)}$}

Let $w_1(x)$ be a weight function which vanishes outside the interval
$x\in[\nu R,R]$, and let $w_1(\nu R)=w_1(R)=0$. Then, the {\em local}
estimate of the multipole moment $Q^{(n)}$ corresponding to this
weight function is given by
\begin{equation}
  Q_1^{(n)} = \int_{\nu R}^R{\rm d}x\,x^{n+1}\,w_1(x)\,g_{\rm t}(x) +
  {1\over n}\int_{\nu R}^R{\rm d}x\,x^{n+1}
  \eck{n w_1(x)+x w'_1(x)}\,g_{\rm r}(x)\;.
\label{eq:21}
\end{equation}
This equation is applicable to the case when multipole information is
to be determined on the circle $x\le R$ (i.e.\ $\nu=0$) or on the ring
$\nu R\le x\le R$ from shear data measured in the same regions.

\subsubsection{Non-local estimates of $Q^{(n)}$}

We now want to address the following question. Suppose data are
available within the radial interval $\nu R\le x\le R$ only, can we
estimate the multipole moments for $x<\nu R$ and $x>R$? Since
$Q^{(n)}_\varphi$ and $Q^{(n)}_x$ are equivalent estimates of
$Q^{(n)}$, one can combine the two with weights $\alpha$ and
$(1-\alpha)$, respectively, to obtain
\begin{eqnarray}
  Q^{(n)} = \alpha Q^{(n)}_\varphi +(1-\alpha) Q^{(n)}_x &=&
  \int_0^\infty{\rm d}x\,
  \eck{(2\alpha-1)x^{n+1}\,w(x)+2(1-\alpha)\,W(x)}\,g_{\rm t}(x)
  \nonumber\\&+&
  \int_0^\infty{\rm d}x\,
  \left[\alpha x^{n+1}\,w(x)+{\alpha\over n}x^{n+2}\,w'(x)
  -n(1-\alpha)W(x)\right]\,g_{\rm r}(x)\;.\nonumber\\
\label{eq:22}
\end{eqnarray}
In the light of this equation, we ask whether we can choose the weight
$\alpha$ and the weight function $w(x)$ such that the integrands in
(\ref{eq:22}) vanish outside $[\nu R,R]$, but with $w(x)$ not
identically zero outside that interval. If that were possible, we
could deduce some information on the multipoles of the deflector in
those regions where no shear information is available.

The requirement that the brackets in the two integrals of (\ref{eq:22})
vanish leads to two solutions for the weight function $w(x)$ and the
corresponding weights $\alpha$,
\begin{equation}
  w_+(x) = {\rm const.}\quad\hbox{with}\quad
  \alpha_+ = {n\over2(n+1)}\quad;\quad
  w_-(x) \propto x^{-2n}\quad\hbox{with}\quad
  \alpha_- = {n\over2(n-1)}\;,
\label{eq:23}
\end{equation}
where the latter is valid only for $n\ge3$.

We first consider the multipole moment within the circle $x\le\nu R$,
i.e.\ interior to the region of the data. Thus, we search for a weight
function $w_2(x)$ which vanishes for $x>R$, but which is finite for
$x<\nu R$. Hence, $w_2(x)$ must be of the form (\ref{eq:23}) for
$x<\nu R$. Since $w_-(x)$ does not satisfy the conditions
(\ref{eq:14}) and (\ref{eq:18}), $w_2(x)={\rm const.}$ for $x\le\nu
R$, and we choose $w_2(x)=1$ for $x\le\nu R$ without loss of
generality. In order for $W_2(x)$ to vanish for $x>R$, we further
require
\begin{equation}
  \int_0^R{\rm d} x\; x^{n+1}\,w_2(x)=0\;.
\label{eq:25}
\end{equation}
Hence, the `internal' multipole moment $Q^{(n)}_2$ corresponding to
the weight function $w_2$ with the aforementioned properties becomes
\begin{equation}
  Q^{(n)}_2 = {1\over 2(1+n)}\,\int_{\nu R}^R{\rm d}x\,
  \bigg\{\big[(4+2n)W_2(x) - 2x^{n+1}w_2(x)\big]\,g_{\rm t}(x) +
  \big[nx^{n+1}w_2(x) + x^{n+2}w'_2(x) - (2+n)nW_2(x)\big]\,
  g_{\rm r}(x)\bigg\}\;,
\label{eq:26}
\end{equation}
where we have inserted the weight $\alpha_+$ from (\ref{eq:23}).

Next, we consider the multipole moment outside the circle $x=\nu R$,
i.e.\ exterior to the region of the data. We therefore require a
weight function which vanishes for $x<\nu R$, but is non-zero for
$x>R$. This weight function, called $w_3(x)$, needs to satisfy the
condition -- see (\ref{eq:14}) and (\ref{eq:18}) --
\begin{equation}
  \int_{\nu R}^\infty{\rm d}x\,x^{n+1}\,w_3(x) = 0\;,
\label{eq:27}
\end{equation}
and if it behaves like $w(x)\propto x^{-2n}$ for $x>R$, then the
multipole moments become, for $n\ge 3$,
\begin{equation}
  Q_3^{(n)} = {1\over 2(n-1)}\int_{\nu R}^R{\rm d}x\,
  \bigg\{\big[2x^{n+1}w_3(x) + (2n-4) W_3(x)\big]\,g_{\rm t}(x) +
  \big[n x^{n+1}w_3(x) + x^{n+2}w'_3-n(n-2)W_3(x)\big]\,
  g_{\rm r}(x)\bigg\}\;.
\label{eq:28}
\end{equation}
This equation is not valid for $n=2$ because then $w\propto x^{-4}$
for $x>R$, and $W(x)$ becomes logarithmic. We have not been able to
find a relation for the {\em external} quadrupole moment. We show in
Appendix B that the shear in the annulus caused by matter not in the
annulus is completely determined by the set of multipole moments of
the matter inside the annulus (i.e., at $x<\nu R$) and the multipoles
of order $n\ge 2$ of the matter outside the annulus.

Of course, the different local and non-local expressions for the mass
multipoles can be combined appropriately. As a simple example,
consider the case that we want to infer $Q^{(n)}$ for $0\le x\le
R$. We then need to combine the {\em internal} multipole moment
(\ref{eq:26}) with the {\em local} multipole moment
(\ref{eq:21}). Thus suppose we have a weight function $w(x)$ which is
constant for $0\le x\le\nu R$ and smoothly falls off to zero for $x\to
R$. Then, $w(x)$ can be decomposed into $w(x)=w_1(x)+w_2(x)$, where
$w_1(x)$ vanishes for $x<\nu R$ and $x>R$, and $w_2(x)$ is constant
for $x<\nu R$, vanishes for $x>R$, and satisfies the constraint
(\ref{eq:25}). Since the aperture multipoles $Q^{(n)}$ are linear in
the weight function $w(x)$, different expressions for $Q^{(n)}$ can be
appropriately superposed. To use a weight function of the form just
discussed, one would have to insert $w_1(x)$ into (\ref{eq:21}) to
obtain $Q_1^{(n)}$ and $w_2(x)=w(x)-w_1(x)$ into (\ref{eq:26}) to
obtain $Q_2^{(n)}$, and the multipole moment would be the sum of the
two, $Q^{(n)}=Q_1^{(n)}+Q_2^{(n)}$.

\section{Multipole signal-to-noise ratio}

The aperture multipole moments calculated in the previous section are
all of the form
\begin{equation}
  Q^{(n)} = \int_{\nu R}^R{\rm d}x\,
  \eck{b_{\rm t}(x)\,g_{\rm t}(x) + b_{\rm r}(x)\,g_{\rm r}(x)}\;.
\label{eq:31}
\end{equation}
With (\ref{eq:20}), we can rewrite (\ref{eq:31}) as
\begin{equation}
  Q^{(n)}(\vec x_0) = \int{\rm d}^2 x\;{\rm e}^{n{\rm i}\varphi}\,
  \rund{{b_{\rm t}(x)\over x}\gamma_{\rm t}(\vec x;\vec x_0) +
  {\rm i} {b_{\rm r}(x)\over x}\gamma_{\rm r}(\vec x;\vec x_0)
  }\;,
\label{eq:32}
\end{equation}
where we have re-introduced the explicit dependence on the center
$\vec x_0$ of the aperture.

In order to apply this equation to data, one needs to approximate the
integral by a sum over individual image ellipticities. Let
$\epsilon_i$ be the complex ellipticity of the $i$-th galaxy image at
position $\vec x_i$. As in Schneider (1996b), the modulus of the
ellipticity for an elliptical image is defined as
$|\epsilon|=(1-r)/(1+r)$, where $r$ is the axis ratio, and the phase
is twice the angle enclosed by the major axis and the positive
$x_1$-axis. In the case of weak lensing, $\kappa\ll 1$, $\epsilon_i$
is an unbiased estimator for the shear $\gamma(\vec x_i)$. Since then
$\gamma$ depends linearly on the distance ratio $D_{\rm ds}/D_{\rm
s}$, the following relations are valid in the case of weak lensing for
a redshift distribution of the sources, with $D_{\rm ds}/D_{\rm s}$
replaced by $\langle D_{\rm ds}/D_{\rm s}\rangle$, the average over
all galaxies used. Since we will mainly be dealing with those regions
of the lenses where $\kappa\ll 1$ (e.g., by choosing the aperture
radii appropriately), we write the discretized version of (\ref{eq:32})
as
\begin{equation}
  Q^{(n)}(\vec x_0) = {1\over\bar n}\sum_{i=1}^N
  {\rm e}^{n{\rm i}\vartheta_i}
  \rund{{b_{\rm t}(y_i)\over y_i}\epsilon_{{\rm t}i} +
  {\rm i}{b_{\rm r}(y_i)\over y_i}\epsilon_{{\rm r}i}}\;,
\label{eq:33}
\end{equation}
where $\bar n$ is the number density of galaxy images in the annulus.
We have introduced polar coordinates ($y_i,\vartheta_i$) such that
\begin{equation}
  \vec x_i = y_i\pmatrix{\cos\vartheta_i\cr
                         \sin\vartheta_i\cr}+\vec x_0\;,
\label{eq:34}
\end{equation}
and the tangential and radial components of the image ellipticity are
defined in analogy with (\ref{eq:11}) as
\begin{equation}
  \epsilon_{{\rm t} i} = 
  -\Re\rund{\epsilon_i\,{\rm e}^{-2{\rm i}\vartheta_i}}\quad;\quad
  \epsilon_{{\rm r} i} =
  -\Im\rund{\epsilon_i\,{\rm e}^{-2{\rm i}\vartheta_i}}\;.
\label{eq:35}
\end{equation}
In the {\em absence} of a lens, the expectation value of $Q^{(n)}$ is
zero, and its dispersion is
\begin{equation}
  \rund{\sigma^{(n)}_{\rm d}}^2 :=
  \left\langle|Q^{(n)}|^2\right\rangle =
  {1\over{\bar n}^2}\,{\sigma_\epsilon^2\over 2}\,
  \sum_{i=1}^N\eck{\rund{b_{\rm t}(y_i)\over y_i}^2 +
  \rund{b_{\rm r}(y_i)\over y_i}^2}\;,
\label{eq:36}
\end{equation}
where we have used that, in the absence of lensing,
\begin{equation}
  \langle\epsilon_{{\rm t} i}\epsilon_{{\rm t} j}\rangle =
  \langle\epsilon_{{\rm r} i}\epsilon_{{\rm r} j}\rangle =
  \delta_{ij}{\sigma_\epsilon^2\over 2}\quad;\quad
  \langle\epsilon_{{\rm t} i}\epsilon_{{\rm r} j}\rangle = 0\;,
\label{eq:37}
\end{equation}
and $\sigma_\epsilon$ is the dispersion of the intrinsic source
ellipticity. The ensemble average of the dispersion is
\begin{equation}
  \rund{\sigma_{\rm c}^{(n)}}^2 = \eck{\prod_{k=1}^N\int_{\nu R}^R
  {2 x_k\;{\rm d} x_k\over (1-\nu^2) R^2}}
  \rund{\sigma^{(n)}_{\rm d}}^2 =
  {\pi\sigma_\epsilon^2\over\bar n}\int_{\nu R}^R{\rm d} y\;
  \rund{b_{\rm t}^2(y)+b_{\rm r}^2(y)\over y}\;.
\label{eq:38}
\end{equation}
We can therefore obtain an approximation for the signal-to-noise ratio
of the multipole moments,
\begin{equation}
  S_{\rm c}^{(n)} = {|Q^{(n)}|\over\sigma_{\rm c}^{(n)}}\;.
\label{eq:39}
\end{equation}

To obtain an estimate for the expected signal-to-noise ratio of the
quadrupole moment of an elliptical mass distribution, we consider a
singular isothermal ellipsoid, with surface-density distribution
\begin{equation}
  \kappa(\vec x) =
  {x_{\rm E}\over 2\sqrt{(1-\eta)^2 x_1^2+(1+\eta)^2 x_2^2}}\;.
\label{eq:40}
\end{equation}
It has an axis ratio of $r=(1-\eta)/(1+\eta)$, and $x_{\rm E}$ is the
characteristic angular scale, which we choose as the Einstein radius
of the corresponding singular isothermal sphere with velocity
dispersion $\sigma_v$,
\begin{equation}
  x_{\rm E} = 4\pi\rund{\sigma_v\over c}^2\,
  \left\langle{D_{\rm ds}\over D_{\rm s}}\right\rangle\;.
\label{eq:41}
\end{equation}
For this mass distribution, the quadrupole moment (\ref{eq:4}) becomes
\begin{equation}
  Q^{(2)} = {x_{\rm E}\,C(\eta)\over 2}\,\int{\rm d}x\,x^{2n}\,w(x)\;,
\label{eq:42}
\end{equation}
where, for moderately small $\eta$, $C(\eta)=\pi\eta+{\cal
O}(\eta^3)$. If we choose (\ref{eq:21}) for the determination of the
quadrupole moment, the signal-to-noise ratio (\ref{eq:39}) reads
\begin{equation}
  S_{\rm c}^{(2)} = {C(\eta)\,x_{\rm E}\sqrt{\bar n}\over 2\sqrt{\pi}
  \sigma_\epsilon}\,{\int{\rm d} x\; x^2\,w(x)\over
  \sqrt{\int{\rm d}x\,x^{5}\rund{2w^2+ xww'+x^2w^{\prime2}/4}}}\;.
\label{eq:43}
\end{equation}

One can now try to find a weight function which maximizes $S_{\rm c}$
for the mass distribution under consideration. From the Cauchy-Schwarz
inequality, it is easy to find that the optimized weight function is
$w(x)\propto x^{-3}$. If this is inserted into (\ref{eq:43}), one sees
that the integration has to be terminated at both ends, say at $x_{\rm
min}$ and $x_{\rm max}$. We then find
\begin{equation}
  S_{\rm c}^{(2)} = {C(\eta)\,x_{\rm E}\,\sqrt{\bar n}\over
  \sqrt{5\pi}\sigma_\epsilon}\,\sqrt{\ln(x_{\rm max}/x_{\rm min})}
  \approx 4.5\,\rund{\eta\over 0.2}\,
  \rund{\sigma_\epsilon\over0.2}^{-1}\,
  \rund{\bar n\over30\,\hbox{arcmin}^{-2}}^{1/2}\,
  \rund{\sigma_v\over1000\,{\rm km\,s^{-1}}}^2\,
  \left\langle{D_{\rm ds}\over D_{\rm s}}\right\rangle\;,
\label{eq:44}
\end{equation}
where is the second step we assumed that $(x_{\rm max}/x_{\rm
min})\sim 10$, and we used the approximation for $C(\eta)$ as given
above.

The optimal weight function cannot be simply chosen as $x^{-3}$, since
we have to satisfy (\ref{eq:14}). For the simulations presented below
we therefore choose a weight function which behaves approximately like
$x^{-3}$ (and, for higher multipole moments, like $x^{-(n+1)}$), but
with appropriate behaviour at $x=\nu R=0$ and $x=R$,
\begin{equation}
  w(x) = {1\over x^{n+1}+x_{\rm min}^{n+1}}
  -{1\over R^{n+1}+x_{\rm min}^{n+1}}
  +{(n+1)R^n(x-R)\over\rund{R^{n+1}+x_{\rm min}^{n+1}}^2}\;.
\label{eq:45}
\end{equation}

\begin{figure}
\begin{center}
  \begin{minipage}[t]{0.5\hsize}
    \epsfxsize=\hsize\epsffile{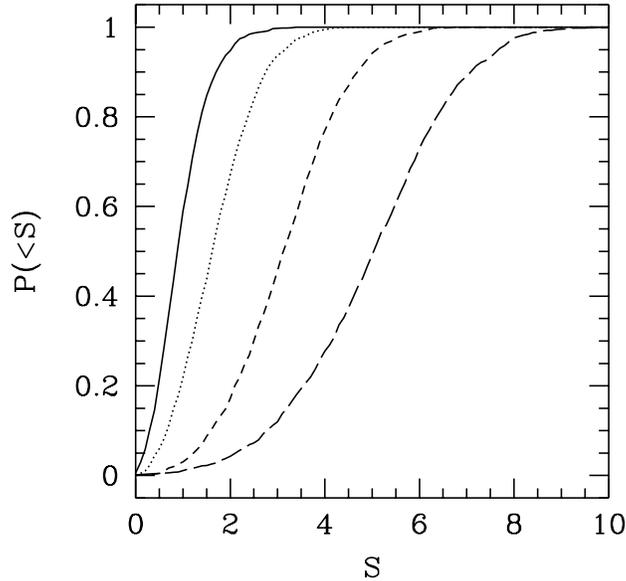}
  \end{minipage}
\end{center}
\caption{The cumulative probability distribution $P(<S)$ for a
  signal-to-noise ratio, as defined in (\ref{eq:48}), being greater
  than $S$.  For these simulations, a galaxy density of 30/arcmin$^2$
  has been assumed, with an intrinsic ellipticity dispersion of
  $\sigma_\epsilon=0.2$ and a mean distance ratio $\langle D_{\rm
  ds}/D_{\rm s}\rangle=0.7$. For the lens, a mass distribution of the
  form (\ref{eq:46}) was assumed, with $\sigma_v=1000\,{\rm
  km\,s^{-1}}$, core size $x_{\rm c}=1.5x_{\rm E}$, and four different
  ellipticity parameters $\eta'=0$ (solid curve), 0.05 (dotted curve),
  0.1 (short-dashed curve) and 0.15 (long-dashed curve)}
\label{fi:2}
\end{figure}

Using this weight function, we have made simulations by assuming a
mass distribution of the form
\begin{equation}
  \kappa(\vec x) = {x_{\rm E}\over2\rund{1-\eta'^2}^2}\,
  \eck{2\rund{1+\eta'^2}x_{\rm c}^2 + |\vec x|^2}\,
  \eck{x_{\rm c}^2+{x_1^2\over (1-\eta')^2}+{x_2^2\over
  (1+\eta')^2}}^{-3/2}\;,
\label{eq:46}
\end{equation}
which corresponds to an elliptical deflection potential of the form
\begin{equation}
  \psi = x_{\rm E}\rund{x_{\rm c}^2 + {x_1^2\over(1-\eta')^2} +
  {x_2^2\over(1+\eta')^2}}^{1/2}\;,
\label{eq:47}
\end{equation}
with axis ratio $(1-\eta')/(1+\eta')$. In the limit of vanishing
ellipticity, $\eta'=0$, and zero core size, $x_{\rm c}=0$,
(\ref{eq:46}) becomes the mass distribution of an isothermal sphere,
with $x_{\rm E}$ given by (\ref{eq:41}). For $|\vec x|\gg x_{\rm c}$,
the isodensity contours have an axis ratio of
$(1-\eta')^3/(1+\eta')^3$, which is 0.74 (0.55, 0.40) for
$\eta'=0.05\;(0.1,0.15)$. Galaxies with $\langle D_{\rm ds}/D_{\rm
s}\rangle=0.7$ and a surface density of $30\,{\rm arcmin}^{-2}$ were
distributed and sheared according to the mass distribution
(\ref{eq:46}). The galaxies were assumed to have a Gaussian ellipticity
distribution with $\sigma_\epsilon=0.2$. In Fig.\ \ref{fi:2} we show
the result of one such simulation, with $\sigma_v=1000\,{\rm
km\,s^{-1}}$, $x_{\rm min}=0.1R$, $R=5'$, $x_{\rm c}=1.5x_{\rm E}$
(i.e., the lens is subcritical), and four different values of
$\eta'$. We have plotted there the cumulative probability for the
signal-to-noise ratio
\begin{equation}
  S = {|Q^{(2)}|\over\sigma_{\rm d}^{(2)}}\;,
\label{eq:48}
\end{equation}
as given by (\ref{eq:33}) and (\ref{eq:36}). As can be seen from Fig.\
\ref{fi:2}, the typical signal-to-noise ratio of the quadrupole moment
for a lens with $\eta'\ga0.1$ is considerably larger than the median
of the $S$-distribution for a spherical lens. That means that for
these mass distributions the quadrupole moment can be detected at a
statistically significant level, whereas for the distribution with
$\eta'=0.05$, there is a considerable overlap between the probability
distribution of the elliptical lens and the corresponding spherical
mass distribution.

\section{Application to numerical cluster simulations}

Having seen in the previous section that the aperture quadrupole from
gravitational shear can significantly be detected for moderately
elliptical idealized lens mass distributions, we proceed here to study
the aperture quadrupole for numerically simulated clusters. From a
large sample of gas-dynamical cluster simulations performed within the
standard CDM cosmogony (for a detailed description of the sample see
Bartelmann \& Steinmetz 1996), we select a subsample of 13 clusters,
all at redshift $z_{\rm d}=0.3$. Since we can project them along three
independent spatial directions, we have in total 39 surface-mass
distributions from these clusters. Assuming sources at redshift
$z_{\rm s}=1$, their maximum convergence values range between
$0.27\le\kappa_{\rm max}\le1.40$, with the median at $\kappa_{\rm
max}\simeq0.73$, and one third of them are critical lenses, i.e., they
produce critical curves.

For each of the 39 simulated cluster fields, we simulate 200 source
galaxy distributions with a galaxy density of $40\,{\rm
arcmin}^{-2}$. Since the fields have a side length of $5'$, the
average number of source galaxies per cluster field is $10^3$. Their
positions are distributed randomly, and they are assigned an
ellipticity drawn from a Gaussian distribution with dispersion
$\sigma_\epsilon=0.15$.

We then determine from each simulation the aperture quadrupole
estimate $Q^{(2)}$ according to (\ref{eq:33}) and the aperture
signal-to-noise ratio according to (\ref{eq:48}), using a weight
function $w(x)$ of the form given in (\ref{eq:45}). There,
$R=2\farcm5$, and we choose $x_{\rm min}=1'$. Since $w(x)$ has the
dimension (length)$^{-3}$, the quadrupole moment has the dimension
(length), and we give it in units of arc minutes in the following
figures.

Figure\ \ref{fi:3} shows for each cluster in the sample the fraction
of the 200 lensing simulations for which the aperture quadrupole is
determined with a signal-to-noise ratio $(S/N)$ larger than some
threshold $(S/N)_0$, as a function of the modulus of the true cluster
quadrupole as defined in (\ref{eq:4}). The difference between the four
panels in the figure is the signal-to-noise threshold $(S/N)_0$. In
panels (a,b,c,d) we have chosen, respectively,
$(S/N)_0=\{2,3,5,10\}$. All panels show that the significance of the
aperture quadrupole measurement increases with increasing intrinsic
quadrupole $|Q^{(2)}_{\rm true}|$, as expected. While the aperture
quadrupole can be determined with $(S/N)>2$ for all clusters with
$|Q^{(2)}_{\rm true}|\ga0\farcm05$ (panel a), $|Q^{(2)}_{\rm
true}\ga0\farcm1$ is required for $(S/N)>5$ (panel c), and only a
small fraction of clusters allows to determine the aperture quadrupole
with $(S/N)>10$ (panel d). For further information, the different
symbols denote the maximum convergence value $\kappa_{\rm max}$ of the
clusters. We use triangles (squares, hexagons) for $\kappa_{\rm
max}<0.7$ ($0.7\le\kappa_{\rm max}<0.83$, $0.83\le\kappa_{\rm max}$),
where the intervals are chosen such that they encompass one third of
the cluster sample each. The distribution of these symbols along the
curves shows that the aperture quadrupole can only be determined
significantly if $\kappa_{\rm max}$ is not too small.

\begin{figure}
\begin{center}
  \begin{minipage}[t]{0.8\hsize}
    \epsfxsize=\hsize\epsffile{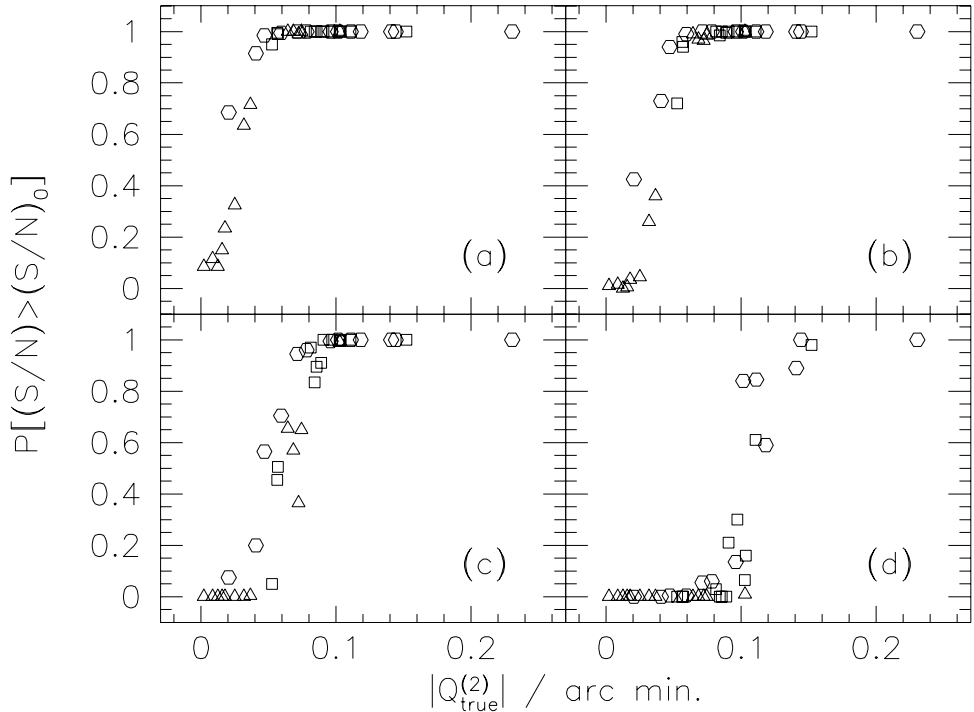}
  \end{minipage}
\end{center}
\caption{For each cluster in the sample with intrinsic quadrupole
  $|Q^{(2)}_{\rm true}|$ as defined in (\protect\ref{eq:4}), the
  fraction of the 200 lensing simulations is shown for which the
  aperture quadrupole is determined with signal-to-noise ratio $(S/N)$
  larger than some limit $(S/N)_0$, where $(S/N)_0=\{2,3,5,10\}$ in
  panels (a,b,c,d), respectively. The triangles (squares, hexagons)
  distinguish between clusters with low (medium, high) maximum
  convergence $\kappa_{\rm max}$ (see text for details). The curves
  show that the aperture quadrupole can be determined with $(S/N)>2$
  for all clusters with $|Q^{(2)}_{\rm true}|\protect\ga0\farcm05$ and
  with $(S/N)>5$ for all clusters with $|Q^{(2)}_{\rm
  true}|\protect\ga0\farcm1$, while only few clusters allow to
  determine the aperture quadrupole with $(S/N)>10$.}
\label{fi:3}
\end{figure}

\begin{figure}
\begin{center}
  \begin{minipage}[t]{0.8\hsize}
    \epsfxsize=\hsize\epsffile{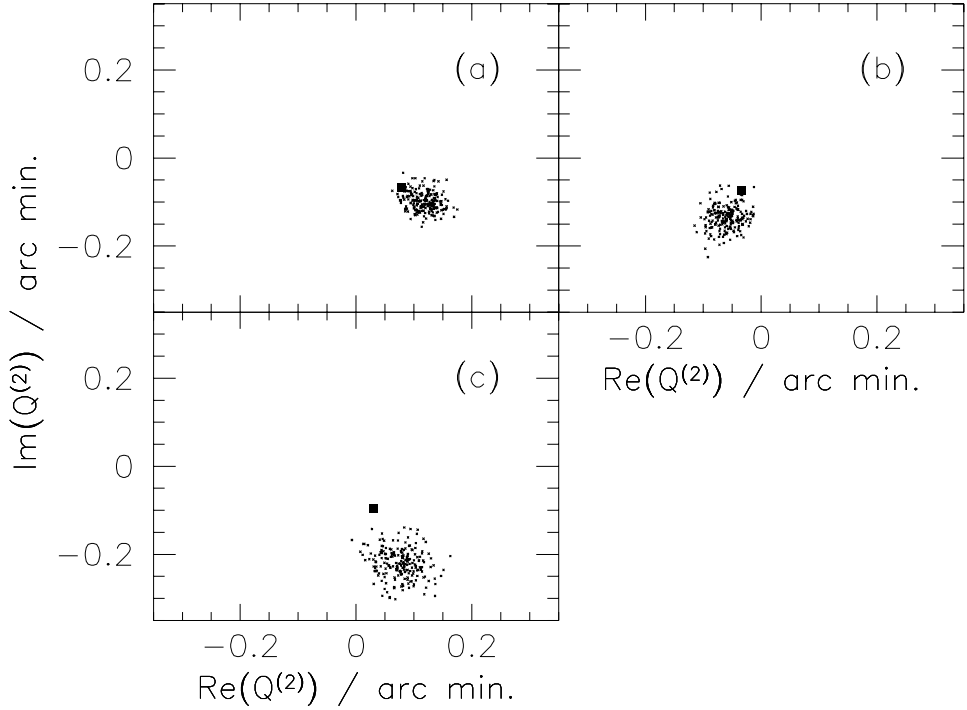}
  \end{minipage}
\end{center}
\caption{Distribution of aperture quadrupole estimates as obtained
  from (\protect\ref{eq:33}) with simulated galaxy ellipticities
  (crosses), together with the true aperture quadrupole (filled
  squares), for three cluster models with approximately equal true
  quadrupole but different maximum convergence values $\kappa_{\rm
  max}$. In panels (a,b,c), $\kappa_{\rm max}=\{0.6,0.7,1.0\}$. The
  figure shows that the measured aperture quadrupole overestimates the
  true one because image ellipticities measure
  $\gamma(1-\kappa)^{-1}>\gamma$ rather than $\gamma$ only.}
\label{fi:4}
\end{figure}

While the shear $\gamma$ enters into the formula for the aperture
multipoles (\ref{eq:13}), the distortion of background objects measures
the reduced shear, which is a combination of shear and convergence,
$g\equiv\gamma(1-\kappa)^{-1}$, rather than the shear. In the case of
weak lensing, $g\approx\gamma$, and then the measured galaxy
ellipticities provide an unbiased estimate for $\gamma$. If lensing is
not weak, however, the observed aperture multipole (\ref{eq:33}) is a
biased estimate of the true multipole. Since $\kappa>0$, $Q^{(n)}$ as
determined from (\ref{eq:33}) overestimates the true aperture
multipole. This is demonstrated in Fig.\ \ref{fi:4}, where we show the
distribution of aperture quadrupole measurements relative to the true
aperture quadrupole for three clusters with approximately equal true
quadrupole, but with different maximum convergence values $\kappa_{\rm
max}$.

As a test, we have verified that the systematic bias in the quadrupole
measurement disappears when we distort the galaxy images with the
shear alone rather than with the reduced shear. Apart from the bias,
the distribution of points in Fig.\ \ref{fi:4} shows that the scatter
in the quadrupole measurements is fairly low.

\section{Discussion}

In this paper we have investigated the possibility to derive multipole
moments of the projected mass distribution of clusters with
weak-lensing techniques. Following the same ideas underlying the
aperture mass measures (Kaiser 1995, Kaiser et al.\ 1994, Schneider
1996b), we have derived several expressions for the mass multipoles of
clusters in terms of the shear distribution in a circle or an annulus
around the cluster center. The shear distribution is observationally
accessible through the ellipticities of faint background galaxies. In
contrast to the aperture masses, the weight functions used for the
multipoles do not need to be of zero total weight, because the
additive constant in the mass reconstruction arising from the
mass-sheet degeneracy is irrelevant for multipoles. 

Our different expressions for the multipole moments can be adapted to
a variety of observational situations:

\begin{itemize}

\item[--] Expression (\ref{eq:21}) permits to determine the multipole
  in a circle or a ring from data given in the same region.
\item[--] Expression (\ref{eq:26}) allows to determine the multipole
  inside a circle from the shear data in an annulus which excludes the
  central region of the circle. This is relevant in cases when the
  central part of a cluster is dominated by bright cluster galaxies
  which outshine the faint background galaxies.
\item[--] Expression (\ref{eq:28}) yields the multipole in a ring as
  before, but also outside the region where data are given. Hence, the
  multipole of a cluster can be determined even if the shear data do
  not cover the entire cluster.
\item[--] As discussed at the end of Sect.\ 2, these different
  expressions can be appropriately combined to match specific
  observational circumstances.

\end{itemize}

Analytic estimates and numerical experiments using simulated cluster
mass distributions demonstrate that the quadrupole of massive clusters
can be detected in a statistically significant way. As an aside, we
have generalized the mass aperture relation to apertures of arbitrary
shape, both for self-similar curves of constant weight (see also
Squires \& Kaiser 1996) and for arbitrary curves of constant weight.

The weight function $w(x)$ used in our numerical studies might not be
the most appropriate choice. In particular, this weight function, when
applied to the multipole formula (\ref{eq:13}), uses shear information
from the entire circle $|\vec x|\le R$ and thus does not punch out a
possible region of strong lensing in the clusters. Therefore, the
measured image ellipticities are not an unbiased estimator of the
shear in the inner part of the cluster, which systematically biases
the aperture quadrupole moments high, as seen in Fig.\
\ref{fi:4}. However, as explained at the end of Sect.\ 2, one can
construct weight functions which avoid this problem and which still
can be optimized for the detection of multipole moments of
quasi-isothermal mass distributions.

Whereas our numerical results on clusters drawn from gas-dynamical
large-scale structure simulations are encouraging and give us faith in
the usefulness of the aperture multipole measures, they should be
considered as a preliminary application only. The cosmological
application of our new approach to quantifying the asymmetry of
cluster mass distributions will require the generation of a sample of
clusters simulated in different cosmological models. The comparison of
the aperture multipole moments of clusters formed in different
cosmologies will then show whether they are sufficiently powerful
indicators of the cosmic density parameter $\Omega$, and how our new
method compares with the one suggested by Wilson et al.\ (1996). To
compare clusters drawn from different cosmological simulations, not
the quadrupole moments themselves should be compared because they
depend on both the clusters mass (or velocity dispersion) and the
degree of asymmetry, but normalized quadrupole moments. As such, the
ratio of the quadrupole moment and the aperture mass seems to be an
appropriate dimensionless measure of the degree of asphericity of a
cluster. A detailed study of these issues, together with a comparison
with the Wilson et al.\ method, will be the subject of a forthcoming
study.

\appendix

\section{Equivalence of $Q^{(n)}_\varphi(x)$ and $Q^{(n)}_x(x)$}

We will now explicitly show the equivalence of (\ref{eq:13}) and
(\ref{eq:17}), starting from a general multipole expansion of the
deflection potential. Choosing coordinates such that $\vec x_0=\vec
0$, we can write
\begin{equation}
  \psi(\vec x) =
  \sum_{m=0}^\infty f_m(x)\,\cos(m\varphi+\varphi_m)\;;
\label{eq:49}
\end{equation}
the function $f_m(x)$ are assumed to have continuous second
derivatives. Using (\ref{eq:8}), (\ref{eq:9}), and (\ref{eq:11}), we can
then write $\gamma_1$, $\gamma_2$, and $\kappa$, and hence also
$\gamma_{\rm t}$, and $\gamma_{\rm r}$ in terms of the functions
$f_m(x)$,
\begin{eqnarray}
  \gamma_{\rm t} &=& -{1\over 2}
  \sum_{m=0}^\infty\cos(m\varphi+\varphi_m)
  \rund{f_m''(x)-{f_m'(x)\over x}+m^2{f_m(x)\over x^2}}\;,
  \nonumber\\
  \gamma_{\rm r} &=& \sum_{m=0}^\infty m\,
  \sin(m\varphi+\varphi_m)\rund{{f_m'(x)\over x} 
  -{f_m(x)\over x^2}}\;.\nonumber\\
\label{eq:50}
\end{eqnarray}

Using further the relations
\begin{equation}
  \int_0^{2\pi}{\rm d}\varphi\,{\rm e}^{n{\rm i}\varphi}\,
  \cos(m\varphi+\varphi_m) =
  \pi\,{\rm e}^{-{\rm i}\varphi_m}\,\delta_{mn}\quad,\quad
  \int_0^{2\pi}{\rm d}\varphi\,{\rm e}^{n{\rm i}\varphi}\,
  \sin(m\varphi+\varphi_m) =
  {\rm i}\pi\,{\rm e}^{-{\rm i}\varphi_m}\,\delta_{mn}\;,
\label{eq:51}
\end{equation}
we find from (\ref{eq:6}) and (\ref{eq:8}) that
\begin{equation}
  Q^{(n)} = {\pi\over 2}\,{\rm e}^{-{\rm i}\varphi_n}
  \int_0^\infty{\rm d}x\, x^{n+1}\,w(x)\,
  \rund{f_n''(x)+{f_n'(x)\over x}-n^2{f_n(x)\over x^2}}\;.
\label{eq:52}
\end{equation}
From (\ref{eq:13}), we find with the same procedure
\begin{eqnarray}
  Q^{(n)}_\varphi = {\pi\over 2}\,{\rm e}^{-{\rm i}\varphi_n}
  &\times&
  \biggl[\int_0^\infty{\rm d}x\, x^{n+1}\,w(x)
  \rund{{f_n'(x)\over x}-n^2{f_n(x)\over x^2}-f_n''(x)}
  \nonumber\\ &-&
  2\int_0^\infty{\rm d}x\,
  \rund{{{\rm d}\eck{x^{n+2}\,w(x)}\over{\rm d} x}-2x^{n+1}\,w(x)}
  \rund{{f_n'(x)\over x}-{f_n(x)\over x^2}}\biggr]\;.
  \nonumber\\
\label{eq:53}
\end{eqnarray}
Integrating the first term of the second integral in (\ref{eq:53}) by
parts and collecting terms, one immediately sees that (\ref{eq:53})
equals (\ref{eq:52}), or that $Q^{(n)}_\varphi =Q^{(n)}$. Similarly, from
(\ref{eq:17}), one has
\begin{equation}
  Q^{(n)}_x = {\pi\over 2}\,{\rm e}^{-{\rm i}\varphi_n}\bigg[
  \int_0^\infty{\rm d}x\,\eck{2W(x)-x^{n+1}\,w(x)}\,
  \rund{{f_n'(x)\over x}-n^2{f_n(x)\over x^2}-f_n''(x)}
  +2n^2\int_0^\infty{\rm d}x\, W(x)\rund{{f_n'(x)\over x}-
  {f_n(x)\over x^2}}\bigg]\;.
\label{eq:54}
\end{equation}
Integrating by parts, using (\ref{eq:16}), and ordering terms yields
that (\ref{eq:54}) also agrees with (\ref{eq:52}), i.e., $Q^{(n)}_x
=Q^{(n)}$.

\section{Influence of matter inside and outside the annulus}

We show here how the shear in the annulus $\nu R\le|\vec x|\le R$ is
affected by matter inside and outside the annulus. In terms of the
surface mass density, the shear is given by
\begin{equation}
  \gamma(\vec x) = {1\over\pi}\int{\rm d}^2y\,
  {\cal D}(\vec x-\vec y)\,\kappa(\vec y)\quad\hbox{with}\quad
  {\cal D}(\vec z) = {-1\over (Z^*)^2}\;,
\label{eq:55}
\end{equation}
where complex notation was used, $Z=z_1+{\rm i}z_2$. For a point $\vec
x$ in the annulus, $\nu R\le|\vec x|\le R$, we then define the {\em
inner} and the {\em outer} shear by
\begin{equation}
  \gamma_{\rm in}(\vec x) = {-1\over\pi}
  \int_0^{\nu R}{\rm d}y\,y\int_0^{2\pi}{\rm d}\vartheta
  {\kappa(\vec y)\over\rund{X^*-y{\rm e}^{-{\rm i}\vartheta}}^2}
  \quad,\quad
  \gamma_{\rm out}(\vec x) = {-1\over\pi}\int_R^\infty{\rm d}y\,y
  \int_0^{2\pi}{\rm d}\vartheta
  {\kappa(\vec y)\over\rund{X^*-y{\rm e}^{-{\rm i}\vartheta}}^2}\;.
\label{eq:56}
\end{equation}
The calculation is performed by first decomposing $\kappa$ in Fourier
modes,
\begin{equation}
  \kappa(\vec y)=\sum_{n=0}^\infty\kappa_n(y)\,
  \cos(n\vartheta+\vartheta_n)\;,
\label{eq:57}
\end{equation}
and then inserting this expression into (\ref{eq:56}). The resulting
$\vartheta$-integrals are evaluated by transforming the integration
variable to $u={\rm e}^{{\rm i}\vartheta}$, ${\rm d}\vartheta=-{\rm
i}\,{\rm d} u/u$, and using the residue theorem. For the inner and
outer shear, we thus obtain
\begin{eqnarray}
  \gamma_{\rm in}(\vec x) = &-&
  {2\over\rund{X^*}^2}\int_0^{\nu R}{\rm d} y\;y\,\kappa_0(y) -
  \sum_{n=1}^\infty
  {(n+1)\,{\rm e}^{{\rm i}\vartheta_n}\over\rund{X^*}^{(n+2)}}
  \int_0^{\nu R}{\rm d} y\;y^{(n+1)}\,\kappa_n(y)\;,\nonumber\\
  \gamma_{\rm out}(\vec x) = &-&
  \sum_{n=2}^\infty{\rm e}^{-{\rm i}\vartheta_n}(n-1)
  \rund{X^*}^{(n-2)}
  \int_R^\infty{\rm d} y\;y^{(1-n)}\,\kappa_n(y)\;.\nonumber\\
\label{eq:58}
\end{eqnarray}
Defining then the {\em inner} and {\em outer} multipole moments
analogously by
\begin{eqnarray}
  Q^{(n)}_{\rm in} &:=& \int_0^{\nu R}{\rm d}y\,y^{(n+1)}
  \int_0^{2\pi}{\rm d}\vartheta\,{\rm e}^{n{\rm i}\vartheta}\,
  \kappa(\vec y) = \pi{\rm e}^{-{\rm i}\vartheta_n}
  \int_0^{\nu R}{\rm d} y\;y^{(n+1)}\,\kappa_n(y)\;,
  \nonumber\\
  Q^{(n)}_{\rm out} &:=& \int_R^\infty{\rm d} y\;y^{(1-n)}
  \int_0^{2\pi}{\rm d}\vartheta\,{\rm e}^{n{\rm i}\vartheta}\,
  \kappa(\vec y) = \pi{\rm e}^{-{\rm i}\vartheta_n}
  \int_R^\infty{\rm d}y\,y^{(1-n)}\,\kappa_n(y)\;,\nonumber\\
\label{eq:59}
\end{eqnarray}
we can write the inner and the outer shear as
\begin{equation}
  \gamma_{\rm in}^*(\vec x) = -{1\over\pi}
  \rund{2 Q_{\rm in}^{(n)}\over X^2}
  +\sum_{n=1}^\infty {(n+1) Q_{\rm in}^{(n)}\over X^{(n+2)}}
  \quad,\quad
  \gamma_{\rm out}(\vec x) = -{1\over\pi}\sum_{n=2}^\infty
  (n-1)\,\rund{X^*}^{(n-2)}\,Q_{\rm out}^{(n)}\;.
\label{eq:60}
\end{equation}
We thus see that the shear in the annulus caused by the matter outside
the annulus is fully described by the multipole moments of this matter
distribution. The shear is affected by all multipoles of the matter
inside the annulus, but only affected by the quadrupole and
higher-order terms outside the annulus. In particular, the outer
quadrupole moment causes a constant shear over the annulus.

\section{Aperture masses for arbitrary aperture shapes}

We show here that the aperture mass measure (\ref{eq:19}) can be
generalized to apertures of arbitrary shape (in fact, also the
aperture multipoles can be generalized, but they will probably be of
less use). As before, we define the aperture mass as
\begin{equation}
  m(\vec x_0) = \int{\rm d}^2y\,
  w(\vec y-\vec x_0)\,\kappa(\vec y)\;,
\label{eq:61}
\end{equation}
where $w$ is a weight function. 

\subsection{Self-similar curves}

Let the weight function be constant on self-similar concentric curves.
Then define a new coordinate system by choosing a closed curve $\vec
c(\lambda)$, $\lambda\in I$, and setting
\begin{equation}
  \vec y = \vec x_0+ b\vec c(\lambda)\;;
\label{eq:62}
\end{equation}
then the mapping between $\vec y$ and $(b,\lambda)$ is one-to-one
(except at $b=0$) iff $\vec c\times\dot{\vec c}\ne 0$ for all
$\lambda\in I$, where $\vec c\times\dot{\vec c}=c_1\dot c_2 -\dot c_1
c_2$. Without loss of generality, choose $\vec c\times\dot{\vec c}\ge
0$. The Jacobian of the coordinate transformation is
$J=\det[\partial(y_1,y_2)/\partial(b,\lambda)]=b\,\vec
c\times\dot{\vec c}$, and so, if $w$ depends on $b$ only, we find from
(\ref{eq:61}) that
\begin{equation}
  m(\vec x_0) = \int_0^\infty{\rm d} b\;b\,w(b)\oint_I{\rm d}\lambda
  \rund{\vec c\times\dot{\vec c}}\;
  \kappa\rund{\vec x_0+b\vec c(\lambda)}\;.
\label{eq:63}
\end{equation}
We now apply the same strategy that was used for deriving $Q_x$ in the
main text. Integrating (\ref{eq:63}) by parts with respect to $b$ yields
\begin{equation}
  m(\vec x_0) = -\int_0^\infty{\rm d} b\;b\,W(b)
  \oint_I{\rm d}\lambda\rund{\vec c\times\dot{\vec c}}\,
  {\partial\kappa\over\partial b}\rund{\vec x_0+b\vec c(\lambda)}\;,
\label{eq:64}
\end{equation}
where we have defined 
\begin{equation}
  W(b) := {1\over b}\int_0^b{\rm d} b'\;b'\,w(b')\;.
\label{eq:65}
\end{equation}
We have further assumed that
\begin{equation}
  \int_0^\infty{\rm d} b\;b\,w(b) = 0
\label{eq:66}
\end{equation}
to ensure that boundary terms vanish. The partial derivative
$\partial\kappa/\partial b$ can now be transformed to
\begin{eqnarray}
  {\partial\kappa\over\partial b} &=&
  c_1{\partial\kappa\over\partial y_1} +
  c_2{\partial\kappa\over\partial y_2} =
  c_1\rund{ {\partial\gamma_1\over\partial y_1} +
  {\partial\gamma_2\over\partial y_2}} +
  c_2\rund{ {\partial\gamma_2\over\partial y_1} -
  {\partial\gamma_1\over\partial y_2}}\nonumber\\
  &=&
  {1\over\vec c\times\dot{\vec c}}
  \eck{\rund{c_1\dot c_2+\dot c_1 c_2}
  {\partial\gamma_1\over\partial b} +
  \rund{\dot c_2 c_2-\dot c_1 c_1}
  {\partial\gamma_2\over\partial b}} +
  {1\over b\,\vec c\times\dot{\vec c}}
  \eck{-2 c_1 c_2{\partial\gamma_1\over\partial\lambda} +
  \rund{c_1^2-c_2^2}{\partial\gamma_2\over\partial\lambda}}\;.
  \nonumber\\
\label{eq:67}
\end{eqnarray}
After inserting (\ref{eq:67}) into (\ref{eq:64}), the terms containing
partial derivatives with respect to $b$ and $\lambda$ are integrated
by parts with respect to the respective coordinate, making use of
(\ref{eq:66}). After collecting terms, this yields
\begin{equation}
  m(\vec x_0) = \int_0^\infty{\rm d} b\;\eck{2 W(b)-b w(b)}
  \oint_I{\rm d}\lambda\,\eck{-\rund{c_1\dot c_2+\dot c_1 c_2}
  \gamma_1+\rund{c_1\dot c_1-c_2\dot c_2}\gamma_2}\;,
\label{eq:68}
\end{equation}
hence the aperture mass $m$ can be expressed directly in terms of
$\gamma$. This result can be written in a more compact form by
defining the function
\begin{equation}
  q(b) := 2bW(b)-b^2w(b) = 2\int_0^b{\rm d} b'\,b'\,w(b)-b^2 w(b)\;,
\label{eq:69}
\end{equation}
and introducing complex notation, $C(\lambda) = c_1(\lambda)+{\rm i}
c_2(\lambda)$. This yields
\begin{equation}
  m(\vec x_0) = \int{\rm d}^2 y\;{q(b(\vec y))\over b^2(\vec y)}\,
  {\Im\rund{\gamma(\vec y)C^*\dot C^*}\over\Im\rund{C^*\dot C}}\;,
\label{eq:70}
\end{equation}
where $C$ has to be taken at $\lambda(\vec y)$. This generalizes the
aperture mass measure as originally derived by Kaiser (1995), Kaiser
et al.\ (1994) and Schneider (1996b); the generalization to
non-circular apertures was first done in Squires \& Kaiser (1996). As
for circular apertures, choosing $w(b)={\rm const.}$ for $0\le b\le\nu
B$ implies $q(b)=0$ for this interval, so that the mass inside the
curve $b=\nu B$ can be measured from the shear around that region.

\subsection{Arbitrary curves}

We now assume that $w(\vec x)$ is constant on a set of nested curves
of arbitrary shapes; in particular, these curves do not have to be
self-similar. In analogy to (\ref{eq:62}), we define a new coordinate
system by
\begin{equation}
  \vec y = \vec x_0+\hat{\vec c}\rund{\hat b,\hat\lambda}\;,
\label{eq:71}
\end{equation}
where $\hat\lambda$ is a cyclic coordinate, $\hat\lambda\in
I=[0,\lambda_{\rm max}]$, and we assume that $w$ depends only on $\hat
b$. In order for (\ref{eq:71}) to define a new coordinate system, we
require that the Jacobian
\begin{equation}
  \hat J(\hat b,\hat\lambda) =
  {\partial\hat{\vec c}\over\partial\hat b}
  \times{\partial\hat{\vec c}\over\partial\hat\lambda}
\label{eq:72}
\end{equation}
is nonzero everywhere (except at the origin). Without loss of
generality, we take $\hat J>0$. The curves will now be relabeled and
reparameterized as follows. Let
\begin{equation}
  {\cal A}(\hat b) = {1\over 2}\oint{\rm d}\hat\lambda\;\hat{\vec c}
  \times{\partial\hat{\vec c}\over\partial\hat\lambda}
\label{eq:73}
\end{equation}
be the area enclosed by the curve labeled with $\hat b$; then we
define a new `radial' coordinate $b(\hat b)$ through
\begin{equation}
  b = \sqrt{2{\cal A}(\hat b)\over\lambda_{\rm max}}\;,
\label{eq:74}
\end{equation} 
so that 
\begin{equation}
  {{\rm d}b\over{\rm d}\hat b} = 
  {1\over\sqrt{2\lambda_{\rm max}{\cal A}(\hat b)}}
  \oint{\rm d}\hat\lambda\;\hat J\rund{\hat b,\hat\lambda}\;.
\label{eq:75}
\end{equation}
We define a new cyclic parameter $\lambda\rund{b,\hat\lambda}\in I$ by
requiring that the Jacobian of the new transformation be independent
of $\lambda$; in particular, the definition of $b$ allows to set the
new Jacobian equal to $b$. Thus, let $\vec c(b,\lambda)=\hat{\vec
c}\rund{\hat b(b),\hat\lambda(b,\lambda)}$, then
\begin{equation}
  J(b,\lambda) = \vec c\ '\times\dot{\vec c} =
  \eck{{{\rm d}\hat b\over{\rm d} b}
  {\partial\hat{\vec c}\over\partial\hat b} +
  {\partial\hat\lambda\over\partial b}
  {\partial\hat{\vec c}\over\partial\hat\lambda}}\times
  {\partial\hat\lambda\over\partial\lambda}
  {\partial\hat{\vec c}\over\partial\hat\lambda} =
  {{\rm d}\hat b\over{\rm d}b}
  {\partial\hat\lambda\over\partial\lambda}
  \hat J\rund{\hat b,\hat\lambda} = b\;,
\label{eq:76}
\end{equation}
where here and in the following, a prime (dot) denotes partial
derivatives with respect to $b$ ($\lambda$). Equation (\ref{eq:76})
determines $\lambda$: using (\ref{eq:74}) and (\ref{eq:75}), we find
\begin{equation}
  \lambda(b,\hat\lambda) =
  {\lambda_{\rm max}\over\oint{\rm d}\hat\lambda\;
  \hat J\rund{\hat b(b),\hat\lambda}}\int_0^{\hat\lambda}
  {\rm d}\tilde\lambda\;\hat J\rund{\hat b(b),\tilde\lambda }\;.
\label{eq:77}
\end{equation}
These definitions of the new coordinates uniquely define at every
point $\vec y$ the two vector fields $\vec c\ '$ and $\dot{\vec
c}$. We now proceed as in Sect.\ C1 before. (\ref{eq:61}) is transformed
into an integral over the new coordinates $(b,\lambda)$, and then
integrated by parts with respect to $b$,
\begin{equation}
  m(\vec x_0) = -\int_0^\infty{\rm d}b\,b\,W(b)
  \oint_I{\rm d}\lambda\,{\partial\kappa\over\partial b}
  \rund{\vec x_0+\vec c(b,\lambda)}\;,
\label{eq:78}
\end{equation}
with $W(b)$ defined as in (\ref{eq:65}), and the condition (\ref{eq:66}) is
applied to $w(b)$. The derivative of $\kappa$ is expressed in terms of
derivatives of the shear, using the coordinate transform and
(\ref{eq:10}) to obtain
\begin{equation}
  {\partial\kappa\over\partial b} = {1\over b}\bigg(\Im (C'\dot C)
  {\partial\gamma_1\over\partial b}-\Re(C'\dot C)
  {\partial\gamma_2\over\partial b} -
  \Im(C'^2){\partial\gamma_1\over\partial\lambda} +
  \Re(C'^2){\partial\gamma_2\over\partial\lambda}\bigg)\;,
\label{eq:79}
\end{equation}
where complex notation was used again, $C'=c_1'+{\rm i}
c_2'$. Inserting (\ref{eq:79}) into (\ref{eq:78}), one obtains after
another partial integration
\begin{equation}
  m(\vec x_0) = \int{\rm d}^2 y\bigg\{
  \rund{{W(b)\over b^2}-{w(b)\over b}}
  \Im\eck{\rund{C'\dot C}^*\gamma} +
  {W(b)\over b}\Im\eck{\rund{C'\dot C'-C''\dot C}^*\gamma}
  \bigg\}\;.
\label{eq:80}
\end{equation}
We note that if one specializes $\vec c(b,\lambda)=b\vec k(\lambda)$
to self-similar curves, (\ref{eq:80}) reduces to (\ref{eq:70}) (note that
(\ref{eq:76}) implies that $\vec k\times\dot{\vec k}=1$). If the weight
function is chosen to be constant within a curve $b_0\vec k(\lambda)$,
one can select the family of curves such that $\vec c(b\lambda)=b\vec
k(\lambda)$ for $b\le b_0$, whereas the curves are arbitrary for
$b>b_0$ (except of course that the family of curves is twice
differentiable). Then, the aperture mass (\ref{eq:61}) can be calculated
from the shear outside the curve labeled by $b_0$. One could, for
example, generalize the $\zeta$-statistics introduced by Kaiser (1995)
to an annular region, where the inner boundary is a circle (e.g.,
situated on the cluster center), and the outer boundary is the
boundary of the CCD. One can easily write down a family of curves
which are circles for $b<b_0$, and which smoothly deform into a square
for $b=B$. The only difference to the standard $\zeta$-statistics is
that the vector fields $\vec c\ '$ and $\dot{\vec c}$ have to be
calculated numerically.

\section*{Acknowledgements}

We thank Matthias Steinmetz for numerically simulating the cluster
models used, and Simon White for useful comments. This work was
supported in part by the Sonderforschungsbereich SFB 375-95 of the
Deutsche Forschungsgemeinschaft.

\end{document}